\documentclass[12pt]{article}

\begin{document}

\title{Generalization of the classical Kramers rate for
        non-Markovian open systems out of equilibrium}

\author{ A. O. Bolivar   \vspace{3pc}\\
\vspace{3pc}
Instituto M\'{a}rio Sch\"{o}nberg \\
Ceil\^{a}ndia, 72225-971, Cx. P. 7316, D.F, Brazil. }

\date{\today}

\maketitle

\begin{abstract}
We analyze the behavior of a Brownian particle
moving in a double-well potential. The escape probability of this
particle over the potential barrier from a metastable state
toward another state is known as the Kramers problem. In this
work we generalize   Kramers' rate theory
to the case of an environment always out of 
thermodynamic equilibrium reckoning with non-Markovian
effects. 
\end{abstract}
\vspace{20pc}


\section{Introduction: Metastability and fluctuations}

We consider a Brownian particle immersed in an environment (e.g.,
a fluid) under the influence of an external potential. Due to the
environmental fluctuations the escape rate of this particle over
the barrier separating two metastable states --in a double-well
potential, for instance-- is known as the Kramers problem
\cite{Kramers40,Hanggi90}, even though the rate theory has been already
tackled by van't Hoff and Arrhenius as late as 1880 \cite{Hanggi90}.
For many years this phenomenon has had various applications in physical,
chemical, astronomical, and biological systems
\cite{Kramers40,Hanggi90,Berliner91,Marchesoni2005}. Originally, 
 Kramers \cite{Kramers40}
investigated the Brownian movement in a reservoir at thermodynamic
equilibrium taking into account only Markovian effects. He also 
 worked out a method for calculating the escape
probability from a Fokker--Planck equation (nowadays known as the
Kramers equation) associated with a given set of Langevin equations. Even
then, several generalizations of this Kramers' pioneering work
have arisen in the literature with experimental verifications
\cite{Hanggi90,Berliner91,Bolivar2005,Comment}, e.g., 
in Josephson junction measuring the decay of the
supercurrent. \vspace{1pc}

 In the theory of escape rate non-Markov and/or nonequilibrium features
 are commonly introduced through memory effects contained
 in the friction kernel present in  generalized Langevin equations and
 using either the Fokker--Planck equation found by Adelman and Mazo
 \cite{Adelman76}
 or  a non-Markovian Smoluchowski equation \cite{Hanggi82}, or yet
 using the Fokker--Planck equation in energy picture
 \cite{Carmeli83}. The equilibrium
 Kramers rate using only the non-Markovian generalized Langevin
 equation is investigated in   \cite{Grote80}.
 In nonequilibrium situations  the Kramers theory has been also
 studied in Markovian open systems with oscillating barriers
 \cite{Marchi96}, as
 well as in periodically driven stochastic systems
 \cite{Gammaitoni98}.\vspace{1pc}

It should be remarked that a feature common to {\it all} above
approaches is that the mean value of the stochastic term present
in the Langevin equations is zero. Following a diverse 
way, in the present paper we propose 
a generalization of the Langevin equations and construct the 
respective Fokker-Planck equation. In
this context we evaluate the Kramers escape rate away from the
equilibrium taking into account non-Markovian effects related to
different time scales inherent in the Brownian dynamics. 
As recently pointed out by Pollak and Talkner \cite{Pollak2005}  
these topics are still poorly understood despite their ongoing relevance 
for rate theory.
\vspace{1pc}

Our paper is organized as follows:

I. Introduction: Metastability and fluctuations

II. Generalizing the Langevin approach

III. Our Fokker--Planck equation

IV. Kramers rate: nonequilibrium and non-Markov

V. Summary and perspectives

Appendix A: Derivation of our Fokker--Planck equation [Eq.(12)]

Appendix B: An example

\section{Generalizing the Langevin approach}

As a physical model of a stochastic process we consider a particle with mass $m$
immersed into an environment.  This particle undergoing a Brownian motion 
is characterized by the  stochastic position $X=X(t)$ and the 
stochastic momentum $P=P(t)$, while the environment is specified by a random
variable $\Psi=\Psi(t)$. Such physical quantities could be intertwined 
through the relations
\begin{equation}
X=Q+\Delta Q\quad \quad;\quad \quad 
P=m\frac{d X}{d t}, 
\end{equation}
where $\Delta Q=\alpha b_1(t)\Psi(t)$, $t$ being a parameter, called time, and   
$\alpha$ a dimensional constant such
that $\Delta Q$ has dimension of length. $d/dt$  denotes a differential operator 
acting upon $X$, and  
$b_1(t)$ a time-dependent parameter measuring the strength of the 
environment effects upon the particle. We define it as being
\begin{equation}
b_1=b_1(t)=\int_{0}^{t}
\langle \Psi(t')\Psi(t'')\rangle dt'',
\end{equation}
where the mean  
$$
\langle \Psi(t')\Psi(t'')\rangle=\int \int \int \psi(t')\psi(t'') 
D_{XP\Psi}(x,p,\psi,t) dxdp d\psi=
$$
$$
\int \psi(t')\psi(t'') 
D_{\Psi}(\psi,t)d\psi
$$
 is calculated in terms of the joint probability density function 
$D_{XP\Psi}(x,p,\psi,t)$ or the probability density $D_{\Psi}(\psi,t)$. 
\vspace{1pc}

One assumes the motion
of the Brownian particle moving in an external potential $V(X)$
to be described by the stochastic differential equations 
in phase space ($X,P$), known as Langevin's equations \cite{Langevin1908}, 
\begin{equation}
\frac{dP}{dt}= -\frac{dV}{dX}-\frac{\gamma}{m}P+b_1\Psi
\quad;\quad \frac{dX}{dt}=\frac{P}{m},
\end{equation}
where
$-\gamma P/m$ denotes a (memoryless) frictional force
activating the particle motion.  There $\Psi$ has the statistical 
properties  
\begin{equation}
\langle \Psi(t')\Psi(t'')\rangle =2D^{1/3}\delta(t''-t')\quad \quad;
\quad \quad \langle \Psi \rangle=0,
\end{equation}
making the stochastic process Markovian. $\delta(t''-t')$ is the 
Dirac delta function and  
$D$ is a  constant -- to be determined by 
the physics of the problem -- such that $b_1\Psi=D^{1/3}$ in Eq.(3) 
has in fact dimension of newton.
\vspace{1pc}

It  is important to note that as  the environmental parameter
$b_1(t)$ does vanish, the stochastic quantities $P$
and $X$ reduce to the respective deterministic values $p=mdq/dt$
and $x=q$, provided $D_{XP}(x,p)=\delta(x-q)\delta(p-p')$. 
Physically, that means that the initially open system becomes isolated 
from its environment and turns out to be described by Newton's equations 
\begin{equation}
\frac{dp}{dt}=-\frac{dV(x)}{dx} -\gamma \frac{p}{m} \quad \quad;\quad \quad 
\frac{dx}{dt}=\frac{p}{m}.
\end{equation}
For this reason one says that the Langevin equations (3) are a generalization 
of Newton's equations (5).
\vspace{1pc}

In the literature \cite{Hanggi90}  
the non-Markovian character is
introduced by means of the following statistical properties of
$\Psi$
\begin{equation}
\langle \Psi(t')\Psi(t'')\rangle =(D/t^2_c)^{1/3}e^{-(t''-t')/t_c}
\quad \quad;\quad \quad
\langle \Psi \rangle=0,
\end{equation}
where $t''>t'$ and $t_c$ is the correlation time between the Brownian particle 
and its environment. One takes into account a memory friction kernel in the 
Langevin equations (3): 
\begin{equation}
\frac{dP}{dt}= - \frac{d V}{d X}-
\int_{0}^{t}\beta(t-\tau)\frac{P(\tau)}{m}d\tau+b_1\Psi
\quad;\quad \frac{dX}{dt}=\frac{P}{m}.
\end{equation}
Both the frictional kernel $\beta(t-\tau)$ and the 
fluctuating function $\Psi(t)$ are coupled 
by means of the dissipation-fluctuation theorem \cite{Zwanzig73}
$$
\langle \Psi(t')\Psi(t'')\rangle =\kappa_B T\beta(t-\tau).
$$ 
Physically, such a theorem assures that the Brownian particle will 
always attain the thermal equilibrium of the heat bath 
characterized by Boltzmann's constant $\kappa_B$ and the temperature 
$T$. 
As $\beta(t-\tau)=2\gamma \delta(t-\tau)$ and 
the correlation time $t_c$ tends to zero, i.e., 
$t_c \rightarrow 0$, the expression  
(6) reduces to (4) while (7) reproduces  (3). 
Thereby, the stochastic dynamics (7), along with the statistical 
properties (6), are called the generalized Langevin equations 
\cite{Zwanzig73}.
\vspace{1pc}

In the present  paper 
our purpose is to make another extension of the Langevin approach. 
To begin with, we hold the definition of $X$ in (1) and  generalize 
the stochastic momentum $P=dX/dt$ according to 
\begin{equation}
\bar P=P+\Delta P,
\end{equation}
where $\Delta P=-mb_2(t)\Psi(t)$, with $b_2(t)$ defined as
\begin{equation}
b_2=b_2(t)=\int_{0}^{t}\langle \Psi(t')\rangle dt'.
\end{equation}
Accordingly, the Langevin equations (3) turn out to be written as 
\begin{equation}
\frac{d\bar P}{dt}= -\frac{dV}{dX}-\frac{\gamma}{m}\bar P+b_1\Psi
\quad;\quad \frac{dX}{dt}=\frac{\bar P}{m}+b_2\Psi,
\end{equation}
in phase space $(X,\bar P)$, with 
\begin{equation}
\langle \Psi(t')\Psi(t'')\rangle =(D/t^2_c)^{1/3}e^{-(t''-t')/t_c}\quad \quad;
\quad \quad \langle \Psi \rangle=(C/t_c^2)^{1/3}e^{-t'/t_c}. 
\end{equation}
As the constant $C$ vanishes, 
we recover from (10) the usual Langevin equations (3) as a special case. 
In short, equations in (10), together with (11), are our generalized Langevin equations. 
\vspace{1pc}


\section{Our Fokker--Planck equation}


Considering a Brownian particle in a harmonic potential $V=kx^2/2$, Equations  
(10) and (11) generate the following Fokker--Planck equation in phase space 
$(x,\bar p)$ (for details, see Appendix A)
\begin{equation}
\frac{\partial {\cal F}}{\partial t}=-\frac{\partial (A_x{\cal
F})}{\partial x}- \frac{\partial (A_{\bar p}{\cal F})}{\partial
\bar p}+\frac{A_{xx}}{2} \frac{\partial^2 {\cal F}}{\partial
x^2}+ A_{x\bar p}\frac{\partial^2 {\cal F}}{\partial x\partial
\bar p}
 +\frac{A_{\bar p \bar p}}{2}\frac{\partial^2 {\cal F}}{\partial \bar p^2},
\end{equation}
where
$$
{\cal F}={\cal F}(x,\bar p,t)=\int D_{X\bar P\Psi}(x,\bar
p,\psi,t)d\psi.
$$
The quantities
$$
A_x=(\bar p/m)+ (C^2/t_c)^{1/3}
(e^{-t/t_c}-e^{-2t/t_c}),
$$
and
$$
A_{\bar p}=-kx -(\gamma /m)\bar p+(CD/t_c)^{1/3}
(e^{-t/t_c}-e^{-2t/t_c})
$$
are the drift coefficients, whereas the time-dependent diffusion coefficients are given by
$$
A_{xx}=(C^2D)^{1/3}(1-e^{-t/t_c})^{2},
$$
$$
A_{x\bar p}=(D^2C)^{1/3}(1-e^{-t/t_c})^{2},
$$
and
$$
A_{\bar p \bar p}=D(1-e^{-t/t_c})^{2}.
$$
Combining $A_{xx}$, $A_{x\bar p}$, and $A_{\bar p \bar p}$ we notice that  
they satisfy the relation
\begin{equation}
\sqrt{A_{xx}A_{\bar p \bar p}}=A_{x\bar p}.
\end{equation}
Moreover, on replacing the 
Maxwell--Boltzmann (MB) distribution 
\begin{equation}
{\cal F}(x,\bar p)=\frac{1}{\sqrt{2\pi mk_B T}}
 e^{-(\bar {p}^2/2 m\kappa_B T)}e^{-(kx^2/2\kappa_B T)}
\end{equation}
into our Fokker--Planck equation (12) it is too easy to verify 
that (14) cannot become its solution. 
This means that our stochastic
process, described by (10--12), holds always away   from the thermal
equilibrium. That leads us to think that the physical meaning 
of the relation (13), which is a consequence of our 
assumption $\langle \Psi \rangle\neq 0$ in (11), is connected 
with nonequilibrium characteristics underlying the environment. 
In fact, as $C=0$ the constraint (13) is broken up and our generalized momentum 
$\bar P$ in Eq.(8) becomes equal to $P$. Consequently, Eq.(12) reduces to the
non-Markovian Kramers equation in phase space $(x,p)$
\begin{equation}
\frac{\partial {\cal F}}{\partial t}=-\frac{p}{m}
\frac{\partial {\cal F}}{\partial x}- \frac{\partial }{\partial p}
\left[\ \left(\ -kx- \frac{\gamma}{m} p \right){\cal F}
\right]+ \frac{D(1-e^{-t/t_c})^{2}}{2}\frac{\partial^2 }{\partial
 p^2}{\cal F}.
\end{equation}
In the Markovian steady regime characterized by  
 $t\gg t_c$, or formally 
$t_c \rightarrow 0$, the MB distribution (14) with $\bar p=p$ 
turns out to be a solution to (15), thereby determining the
diffusion coefficient as being equal to $A_{pp}=D=2\gamma
\kappa_{B} T$. 
\vspace{1pc}

On the other hand, inserting
${\cal F}(x,\bar p,t)=f(x,t)\delta(\bar p)$ into (12) 
and taking into account the high friction condition
$$
\gamma \frac{\bar p}{m}=-kx,
$$
obtained from Newton's equations (5) on neglecting inertial effects 
($|d\bar p/dt|\ll|\gamma \bar p/m|$), we
arrive at  the non-Markovian Smoluchowski equation in position space
\begin{equation}
\frac{\partial f(x,t)}{\partial t}=-\frac{1}{\gamma}
\frac{\partial }{\partial x}[{\cal K}(x,t)f(x,t)]+
\frac{A_{xx}}{2}\frac{\partial^2 f(x,t)}{\partial x^2},
\end{equation}
where 
$$
{\cal K}(x,t)=-kx+
\gamma(C^2/t_c)^{1/3}(e^{-t/t_c}-e^{-2t/t_c}).
$$
 Replacing (14) into (16) we obtain 
$A_{xx}=2\kappa_B T/\gamma$ in both
stationary and Markovian regimes.
\vspace{1pc}

Considering $V=0$ and $C=0$ 
from our equation (12)  we derive the 
non-Markovian Rayleigh equation in  $p$-space 
\cite{Kampen81}
\begin{equation} 
\frac{\partial g(p,t)}{\partial t}=\frac{\gamma}{m}
\frac{\partial }{\partial p}[pg(p,t)]+
\frac{D(1-e^{-t/t_c})^2}{2}\frac{\partial^2 }{\partial p^2}g(p,t),
\end{equation}
with 
$$ 
g(p,t)=\int D_{XP}(x,p,t)dx.
$$
\vspace{1pc}

From the mathematical viewpoint we note that    
we can derive the Kramers equation (15), the Smoluchowski equation (16), 
and the Rayleigh equation (17) as special cases of our equation of 
motion (12). Physically, that means that all the physics encapsulated 
into these equations of motion (15), (16), and (17) are in principle 
contained in our Eq.(12). 


\section{Kramers rate: Nonequilibrium and non-Markov}


In order to provide  a physical significance to our 
equation of motion (12) we are going to evaluate the 
Kramers rate.  This problem consists on calculating 
the escape probability of a Brownian particle over 
a potential barrier in a presence of an environment 
away from equilibrium and having non-Markovian features. 
\vspace{1pc}

We follow the Kramers' approach \cite{Kramers40,Kampen81,Risken89,Gardiner85} for 
calculating escape rate.  
We start with the non-Markovian Kramers equation (15)
and derive a  stationary  equation as 
we assume that during a given fixed time interval $t=\Delta \tau$ 
for observing the Brownian particle, the function ${\cal F}$ can be factorised as  
\begin{equation}
{\cal F}(x,p,t)=e^{\gamma t/m}F(x,p)|_{t=\Delta \tau}.
\end{equation}
Inserting (18) into (15) leads to the stationary Kramers equation
\begin{equation}
-\frac{p}{m}
\frac{\partial F}{\partial x}- \frac{\partial }{\partial
p}\left[\ \left(\ -kx- \frac{\gamma}{m} p \right)F
\right]+ \frac{D(1-e^{-\Delta \tau/t_c})}{2}\frac{\partial^2 }{\partial
p^2}{F}=0.
\end{equation}
 \vspace{1pc}
 
After the change of variable  according to 
\begin{equation}
\xi=p-ax
\end{equation}
Eq.(19) turns into 

\begin{equation}
\frac{d^2F}{d\xi^2}=-A\xi\frac{dF}{d\xi},
\end{equation}
 where 

\begin{equation}
A=\frac{2(a-\gamma)}{mD(1-e^{-\Delta \tau/t_c})}.
\end{equation}
Solution to (21) is given by [after using condition $F(\xi=\infty)=1$]
\begin{equation}
F(\xi)= \left(\ \frac{A}{2\pi} \right)^{1/2}
\int_{-\infty}^{\xi}e^{-A\xi^2/2} d\xi,
\end{equation}
provided $A>0$, i.e.,
\begin{equation}
a=\frac{1}{2}\left(\ \gamma + \sqrt{\gamma^2-4mk}\right), \quad k>0.
\end{equation}
 Following Kramers \cite{Kramers40} and using (23) we build up the 
 following function
\begin{equation}
{\cal W}(x, p)=e^{-\beta[( p^2/2m)+V(x)]}F(\xi= p-ax),
\end{equation}
where the parameter $\beta$ has dimension of 1/joule, so that the
exponential in (25) is dimensionless. 
[Notice that on account of the temperature
concept cannot generally be defined in nonequilibrium situations
we could not employ the thermodynamic identity $\beta=1/\kappa_{B}T$].
From (26) and (27) we notice that $\beta$ should be  
restricted to the values $0<\beta<\infty$.
\vspace{1pc}

In a double-well potential
$V(x)$ the barrier top is located at point $x_b$, whereas
the metastable wells are at $x_a$ and $x_c$, with
$x_c>x_a$ such that $V(x_a)=V(x_c)=0$. To find out 
the probability current (flux) over the potential barrier located
at $x=x_b$, i.e.,
$$
j_b=\int_{-\infty}^{+\infty}{\cal W}(x=x_b,p)\frac{p}{m}dp,
$$
we expand $V(x)$ and $\xi$ about $x_b$:
$$
V(x)\approx V(x_b)-(m\omega^{2}_{b}/2)(x-x_b)^2,
$$
and
$$
\xi=p-a(x-x_b).
$$
 $\omega_{b}$ is the oscillation frequency over the barrier.
Consequently, using (25) we find the probability current
\begin{equation}
j_{b}=\frac{1}{\beta}\left(\  \frac{Am}{Am+\beta}   \right)^{1/2}
e^{-\beta V(x_b)}
\end{equation}
with $a$ in (24) and (22) given by
$$
a=\frac{1}{2}\left(\ \gamma + \sqrt{\gamma^2+4m^2\omega^2_b}\right).
$$
\vspace{1pc}

The number of particles $\nu_a$ in the metastable state around $x_a$ can be calculated
with (25) in the  limit $\xi \rightarrow \infty$ as being
\begin{equation}
\nu_a=\int_{-\infty}^{+\infty}\int^{+\infty}_{-\infty}
e^{-\beta[(p^2/2m)+(k_ax^2/2)]}dp dx=\frac{2\pi}{\omega_a \beta}, \quad
k_a=m\omega^2_a.
\end{equation}
\vspace{1pc}

Using (26) and (27) we derive the nonequilibrium and non-Markov  escape rate  
\begin{equation}
\Gamma_{{\mbox {neq}}}=\frac{j_{b}}{\nu_a}=
\frac{\omega_a}{2\pi}
\left(\  \frac{Am}{Am+\beta}   \right)^{1/2}
e^{-\beta V(x_b)}.
\end{equation}
\vspace{1pc}

Physical regimes related to the time scales 
$\Delta \tau$ (observation time) and $t_c$ 
(correlation time):\vspace{1pc}

i) For $\Delta \tau=t_c$, the non-Markovian, non-equilibrium escape rate is given by (28) with
 
$$
A\approx 3\frac{(a-\gamma)}{mD};
$$

ii) while for $\Delta  \tau \ll t_c$ ($t_c \rightarrow \infty$), the system holds 
highly non-Markovian at non-equilibrium situation having the rate (28) with 
$$
A\approx \frac{2t_c(a-\gamma)}{mD\Delta \tau};
$$

iii) for $\Delta \tau  \gg t_c  $ ($t_c \rightarrow 0$) the Markovian regime 
is attained  along with the equilibrium state. In this context, 
$\beta =1/\kappa_B T$ and the diffusion 
constant can be calculated as being $D=2\gamma \kappa_B T$. Thereby, from (28) 
we obtain the well-known Markovian Kramers rate at  thermodynamic equilibrium  
\begin{equation}
\Gamma_{{\mbox {eq}}}=\frac{\omega_a}{2\pi m\omega_b}
\left[\ \sqrt{\frac{\gamma^2}{4}+m^2\omega^2_b} -\frac{\gamma}{2}\right]
e^{- V(x_b)/\kappa_B T}.
\end{equation}
\vspace{1pc}

Now we want to extend Kramers' approach to our Fokker--Planck 
equation (12). To this end, we assume again the function
$$
{\cal F}(x,\bar p,t)=e^{\gamma t/m}F(x,\bar p)
$$
to be a solution to Eq.(12) during a 
fixed time interval $\Delta \tau$ (a time of observation).  As above we  
perform the variable change $\xi=\bar p-ax$ 
and  derive the following ordinary differential equation from (12)
\begin{equation}
\frac{d^2F}{d\xi^2}=-({\cal A}\xi-{\cal B})\frac{dF}{d\xi},
\end{equation}
where
\begin{equation}
{\cal A}=\frac{2(a-\gamma)}{m\left(\ a\sqrt{A_{xx}}+
\sqrt{A_{\bar p \bar p}}\right)^2} \quad;\quad
{\cal B}=
\frac{2b({\Delta \tau})[1+a(C/D)^{1/3}]}
{\left(\ a\sqrt{A_{xx}}+\sqrt{A_{\bar p \bar p}}\right)^2},
\end{equation}
with
$$
b(\Delta \tau)=(CD/t_c)^{1/3}(e^{-\Delta \tau/t_c}-e^{-2\Delta
\tau/t_c}),
$$
$$
A_{xx}=(C^2D)^{1/3}(1-e^{-\Delta \tau/t_c})^{2},
$$
$$
A_{x\bar p}=(D^2C)^{1/3}(1-e^{-\Delta \tau/t_c})^{2},
$$
$$
A_{\bar p \bar p}=D(1-e^{-\Delta \tau/t_c}),
$$

and
$$
a=\frac{1}{2}\left(\ \gamma \pm \sqrt{\gamma^2-4mk}\right).
$$
Solution to (30) is given by [after using condition $F(\xi=\infty)=1$]
\begin{equation}
F(\xi)= \left(\ \frac{{\cal A}}{2\pi}\right)^{1/2}e^{-{\cal B}^2/2{\cal A}}
\int_{-\infty}^{\xi}e^{-({\cal A}/2)\xi^2+{\cal B}\xi}d\xi,
\end{equation}
provided ${\cal A}>0$, i.e.,
\begin{equation}
a=\frac{1}{2}\left(\ \gamma + \sqrt{\gamma^2-4mk}\right), \quad k>0.
\end{equation}
We build up the function
\begin{equation}
{\cal W}(x,\bar p)=e^{-\beta[(\bar p^2/2m)+V(x)]}F(\xi=\bar p-ax).
\end{equation}
\vspace{1pc}

As supposed above, 
let the Brownian particle be to move in a double-well potential
$V(x)$, in which the barrier top is located at point $x_b$, whereas
the metastable wells are at $x_a$ and $x_c$, with
$x_c>x_a$ such that $V(x_a)=V(x_c)=0$. In order to find
the flux over the potential barrier located
at $x=x_b$,
$$
j_b=\int_{-\infty}^{+\infty}{\cal W}(x=x_b,\bar p)(\bar p/m)d\bar p,
$$
we expand $V(x)$ and $\xi$ about $x_b$:
$$
V(x)\approx V(x_b)-(m\omega^{2}_{b}/2)(x-x_b)^2,
$$
and
$$
\xi=\bar p-a(x-x_b).
$$
After inserting (34), with (32) and (33), into $j_b$ we find 
\begin{equation}
j_{b}=\frac{1}{\beta}\left(\ \frac{{\cal A}m}{{\cal A}m+\beta} \right)^{1/2}
e^{-\beta {\cal B}^2/2{\cal A}({\cal A}m+\beta)}e^{-\beta V(x_b)}
\end{equation}
while the number of particles $\nu_a$ in the metastable state around $x_a$ is given by 
\begin{equation}
\nu_a=\int_{-\infty}^{+\infty}\int^{+\infty}_{-\infty}
e^{-\beta[(\bar p^2/2m)+(k_ax^2/2)]}d\bar p dx=\frac{2\pi}{\beta \omega_a}, \quad
k_a=m\omega^2_a.
\end{equation}
\vspace{1pc}

With $j_b$ and $\nu_a$ we obtain then the nonequilibrium and non-Markovian
escape rate in the stationary regime 
\begin{equation}
\Gamma_{{\mbox {neq}}}=\frac{j_b}{\nu_a}=\frac{\omega_a}{2\pi}
\left(\ \frac{{\cal A}m}{{\cal A}m+\beta} \right)^{1/2}
e^{-\beta {\cal B}^2/2{\cal A}({\cal A}m+\beta)}e^{-\beta V(x_b)},
\end{equation}
where ${\cal A}$ and ${\cal B}$ are given by (31) with 
$$
a=\frac{1}{2}\left(\ \gamma + \sqrt{\gamma^2+4m^2\omega^2_b}\right).
$$
On account of the exponential term the performance of the escape rate 
(37) is thoroughly controlled by the nonequilibrium parameter $\beta$.
[As $C=0$, our rate (37) reduces to Eq.(28)].
\vspace{1pc}

In the Markovian limit, $t_c \rightarrow 0$, Eq.(37) leads to   
\begin{equation}
\Gamma_{{\mbox {neq}}}=\frac{\omega_a}{2\pi}
\left(\ \frac{2(a-\gamma)}{2(a-\gamma)+\beta\left(\ a\sqrt{
C^2 D)^{1/3}}+\sqrt{D}\right)^{2}}      \right)^{1/2}
e^{-\beta V(x_b)},
\end{equation}
 whereas 
for $C=0$, $D=2\gamma \kappa_B T$, and  $\beta=(1/\kappa_B T)$ 
our result (37), via (38), reproduces  the Markovian escape rate (29)
found by Kramers for a Brownian particle immersed
in a thermal reservoir at thermodynamic equilibrium.\vspace{1pc}

Comparing our result (38) with (29) we arrive at the relation 

\begin{equation}
\frac{\Gamma_{{\mbox {neq}}}}{\Gamma_{{\mbox {eq}}}}=\frac{2m\omega_b}{\sqrt{2(a-\gamma)}}
\frac{1}{\sqrt{2(a-\gamma)+\beta\left(\ a\sqrt{(C^2 D)^{1/3}}+
\sqrt{D}\right)^{2}     }}
e^{V(x_b)(1-\beta\kappa_B T)/\kappa_B T}.
\end{equation}
As $\beta < 1/\kappa_B T$ our nonequilibrium rate $\Gamma_{{\mbox {neq}}}$ is 
enhanced in comparison with the equilibrium Kramers rate $\Gamma_{{\mbox {eq}}}$. On the 
contrary, for $\beta > 1/\kappa_B T$ we find $\Gamma_{{\mbox {neq}}}<\Gamma_{{\mbox {eq}}}$. 
\vspace{1pc}

Although our escape rate $\Gamma_{{\mbox {neq}}}$ (38) cannot be defined 
for $\beta=0$ we may conjecture  about the  mathematical  
behavior of $\Gamma_{{\mbox {neq}}}/\Gamma_{{\mbox {eq}}}$ as $\beta\rightarrow 0$ 
corresponding to  
the extreme physical situation $\beta \ll 1/\kappa_B T$.  From (39) we obtain 
then the result 
\begin{equation}
\frac{\Gamma_{{\mbox {neq}}}}{\Gamma_{{\mbox {eq}}}}=\frac{2m\omega_b}
{\sqrt{\gamma^2+4m^2\omega^2_b}-\gamma}
e^{V(x_b)/\kappa_B T}
\end{equation}
which leads to  
\begin{equation}
\frac{\Gamma_{{\mbox {neq}}}}{\Gamma_{{\mbox {eq}}}}=\frac{\gamma}
{2m\omega_b}
e^{V(x_b)/\kappa_B T}
\end{equation}
for $\gamma \gg 2m\omega_b$ (or formally $\gamma \rightarrow \infty$), and to  
\begin{equation}
\frac{\Gamma_{{\mbox {neq}}}}{\Gamma_{{\mbox {eq}}}}=
e^{V(x_b)/\kappa_B T}
\end{equation}
for $\gamma \rightarrow 0$ (or physically $\gamma \ll 2m\omega_b$). 
By considering the case $\gamma=2m\omega_b$, from Eq.(40) we derive  
\begin{equation}
\frac{\Gamma_{{\mbox {neq}}}}{\Gamma_{{\mbox {eq}}}}=(1+\sqrt{2})
e^{V(x_b)/\kappa_B T}.
\end{equation}
\vspace{1pc}

By comparing  (37) with (38) we can  
evaluate the influence of the non-Markovian effects on  
escape rates out equilibrium. The exponential term  
$$
e^{-\beta {\cal B}^2/2{\cal A}({\cal A}m+\beta)}
$$
therefore does account for diminishing the probability of escape in the 
non-Markovian regime. 
\vspace{1pc}

Before closing this section we want to emphasize that
our general escape rate (37) has been obtained in the aftermath of 
the hypothesis $\langle \Psi \rangle \neq 0$
assumed in Eqs. (10--12), noticing that 
$t$ is the evolution time of the Brownian particle immersed in a fluid,
$\Delta \tau$ denotes an observation time necessary to detect generally 
nonequilibrium physical 
properties at the stationary state, and $t_c$ the correlation time 
responsible for non-Markovian features.

\section{Summary and perspectives}

In this paper we have investigated the metastability phenomenon 
in the presence of fluctuations. 
In Section II we have obtained the  generalized Langevin equations  
(10) on the basis 
of an extension of the stochastic momentum (8) taking into 
account the nonvanishing average value of the random function $\Psi$.
\vspace{1pc}

In Section III we have  built up a Fokker--Planck equation [Eq.(12)] from 
which we have found out the non-Markovian escape rate away 
from the equilibrium (37). As compared to the equilibrium 
rate our result (39) predicts that the probability of escape may decrease 
or increase in the nonequilibrium regime. The parameter $\beta$ 
controls  the performance of such escape rate.  
 \vspace{1pc}

 Throughout our article we have deemed 
 the stochastic system to hold at a nonequilibrium state even in 
 steady situations. Which of many possible physical mechanisms is 
 responsible for approaching it to the equilibrium state? 
 The Markovian character, i.e., as the correlation 
 time $t_c$ is too tiny in comparison with the 
 observation time $\Delta \tau$, seems to be a strong 
 criterion to attain the equilibrium state, provided in  
 our result (38) the constant $C$ does vanish. It is worth 
 remembering that $C$ has been introduced in Eq.(11) for 
 $\langle \Psi \rangle \neq 0$.
 \vspace{1pc}

 Quantum and nonlinear effects of the 
 potential barrier will be studied in a forthcoming work
 \cite{Bolivar}. 
 We hope thus our present approach could  contribute to the formulation 
 of a general theory of escape rate and 
 stimulate experimental researches 
 in the area of non-Markovian escape rate in systems away from thermal 
 equilibrium.

\section *{Acknowledgments}
The author cordially  thanks Dr. Annibal Figueiredo for the scientific support.
\vspace{35pc}


\section*{Appendix A: Derivation of our Fokker--Planck equation [Eq.(12)]}


In this appendix  we want to show in somewhat details how we could explicitly construct 
the Fokker--Planck equation (12) from the system of stochastic differential 
equations \cite{Bolivar2004} 
\begin{equation}
\frac{d\bar P}{dt}= -\frac{dV}{dX}-\frac{\gamma}{m}\bar P+b_1\Psi
\quad;\quad \frac{dX}{dt}=\frac{\bar P}{m}+b_2\Psi.
\end{equation}
Equations (44) yield the results
\begin{equation}
 \Delta \bar P =- \left(\ \frac{dV}{dX}+\frac{\gamma}{m}\bar P\right)\Delta t+
\int_{t}^{t+\Delta t}
b_1(t')\langle \Psi(t')\rangle dt'
\end{equation}
and
\begin{eqnarray}
 \Delta  X =\frac{\bar P}{m}\Delta t- \left(\ \frac{dV}{dX}+
 \frac{\gamma}{m}\bar P\right)\frac{(\Delta t)^2}{m}+\frac{1}{m}
\int_{t}^{t+\Delta t}\int _{t}^{s}b_1(t')\langle \Psi(t')
\rangle dt'ds+
\nonumber \\
\int_{t}^{t+\Delta t}b_2(t')\langle \Psi(t')\rangle dt'.
\end{eqnarray}
Using $\Delta \bar P=\bar P(t+\Delta t)-\bar P(t)$ and 
$\Delta X=X(t+\Delta t)-X(t)$ we calculate the following quantities

\begin{equation}
A_{\bar p}=\lim_{ \Delta t \rightarrow 0}\frac{\langle \Delta \bar P \rangle}
{\Delta t}=-\frac{dV}{dX}-\frac{\gamma}{m}\bar P+
\lim_{ \Delta t \rightarrow 0}\frac{1}{\Delta t}\int_{t}^{t+\Delta t}
b_1(t')\langle \Phi(t')\rangle dt',
\end{equation}

\begin{eqnarray}
A_{p\bar p}=\lim_{ \Delta t \rightarrow 0}\frac{\langle (\Delta \bar P)^2 \rangle}
{\Delta t}=-2\left(\ \frac{dV}{dX}+\frac{\gamma}{m}\bar P\right)
\lim_{ \Delta t \rightarrow 0}\int_{t}^{t+\Delta t}
b_1(t')\langle \Psi(t')\rangle dt'+
\nonumber \\
\lim_{ \Delta t \rightarrow 0}\frac{1}{\Delta t}
\int_{t}^{t+\Delta t}\int_{t}^{t+\Delta t}
b_1(t')b_1(t'')\langle \Psi(t')\Psi(t'')\rangle dt'dt'',
\end{eqnarray}
\begin{eqnarray}
A_{x}=\lim_{ \Delta t \rightarrow 0}\frac{\langle \Delta X \rangle}
{\Delta t}=\frac{\bar P}{m}+\frac{1}{m}
\lim_{ \Delta t \rightarrow 0}\int_{t}^{t+\Delta t}\int_{t}^{s}
b_1(t')\langle \Psi(t')\rangle dt'ds+
\nonumber \\
\lim_{ \Delta t \rightarrow 0}
\int_{t}^{t+\Delta t}b_2(t')\langle \Psi(t')\rangle dt',
\end{eqnarray}
\begin{equation}
A_{xx}=\lim_{ \Delta t \rightarrow 0}\frac{\langle (\Delta X)^2 \rangle}
{\Delta t}=I_1+I_2+I_3+I_4
\end{equation}
with

\begin{equation}
I_{1}=\frac{2\bar P}{m^2}\int_{t}^{t+\Delta t}\int_{t}^{s}
b_1(t')\langle \Psi(t')\rangle dt'ds,
\end{equation}
\begin{equation}
I_{2}=\frac{2\bar P}{m}\int_{t}^{t+\Delta t}b_2(t')\langle \Psi(t')\rangle dt',
\end{equation}
\begin{equation}
I_{3}=\frac{1}{m^2}\lim_{ \Delta t \rightarrow 0}
\int_{t}^{t+\Delta t}\int_{t}^{t+\Delta t}\int_{t}^{r}\int_{t}^{s}
b_1(t')b_1(t'')\langle \Psi(t')\Psi(t'')\rangle dt'dt''drds,
\end{equation}

\begin{equation}
I_{4}=\frac{2}{m}\lim_{ \Delta t \rightarrow 0}
\int_{t}^{t+\Delta t}\int_{t}^{t+\Delta t}\int_{t}^{s}
b_1(t')b_2(t'')\langle \Psi(t')\Psi(t'')\rangle dt'dt''ds,
\end{equation}

\begin{equation}
I_{5}=\int_{t}^{t+\Delta t}\int_{t}^{t+\Delta t}
b_1(t')b_2(t'')\langle \Psi(t')\Psi(t'')\rangle dt'dt'',
\end{equation}

and 
\begin{equation}
A_{x\bar p}=\lim_{ \Delta t \rightarrow 0}\frac{\langle \Delta X \Delta \bar P\rangle}
{\Delta t}=\xi_1+\xi_2+\xi_3+\xi_4+\xi_5,
\end{equation}
where 
\begin{equation}
\xi_1=\frac{\bar P}{m}\int_{t}^{t+\Delta t}
b_1(t')\langle \Psi(t')\rangle dt',
\end{equation}
\begin{equation}
\xi_2=-\frac{1}{m}\left(\ \frac{dV}{dX}+\frac{\gamma}{m}\bar P\right)
\int_{t}^{t+\Delta t}\int_{t}^{s}
b_1(t')\langle \Psi(t')\rangle dt'ds,
\end{equation}
\begin{equation}
\xi_3=\frac{1}{m}\lim_{ \Delta t \rightarrow 0}\frac{1}{\Delta t}
\int_{t}^{t+\Delta t}\int_{t}^{t+\Delta t}\int_t^s
b_1(t')b_1(t'')\langle \Psi(t')\Psi(t'')\rangle dt'dt''ds,
\end{equation}
\begin{equation}
\xi_4=-\frac{\bar P}{m}\int_{t}^{t+\Delta t}
b_2(t')\langle \Psi(t')\rangle dt'
\end{equation}
\begin{equation}
\xi_5=\lim_{ \Delta t \rightarrow 0}\frac{1}{\Delta t}
\int_{t}^{t+\Delta t}\int_{t}^{t+\Delta t}
b_2(t')b_2(t'')\langle \Psi(t')\Psi(t'')\rangle dt'dt''.
\end{equation}
\vspace{1pc}

After using our definitions 

\begin{equation}
\langle \Psi(t')\Psi(t'')\rangle =(D/t^2_c)^{1/3}e^{-(t''-t')/t_c};
\quad \quad \langle \Psi \rangle=(C/t_c^2)^{1/3}e^{-t'/t_c},
\end{equation}
and
\begin{equation}
b_1=\int_{0}^{t}
\langle \Psi(t')\Psi(t'')\rangle dt''=(Dt_c)^{1/3}(1-e^{-t/t_c}),
\end{equation}
\begin{equation}
b_2=\int_{0}^{t}
\langle \Psi(t')\rangle dt'=(Ct_c)^{1/3}(1-e^{-t/t_c}).
\end{equation}
into (47--50) and (56) we obtain our Fokker--Planck equation (12) with 
the coefficients
\begin{equation}
A_x= (\bar p/m)+ \left(\ \frac{C^2}{t_c} \right)^{1/3}
(e^{-t/t_c}-e^{-2t/t_c}),
\end{equation}
\begin{equation}
A_{\bar p}=-kx -\left(\ \frac{\gamma}{m} \right)\bar p+\left(\ \frac{CD}{t_c} \right)^{1/3}
(e^{-t/t_c}-e^{-2t/t_c}),
\end{equation}
\begin{equation}
A_{xx}=(C^2D)^{1/3}(1-e^{-t/t_c})^{2},
\end{equation}
\begin{equation}
A_{x\bar p}=(D^2C)^{1/3}(1-e^{-t/t_c})^{2},
\end{equation}
\begin{equation}
A_{\bar p \bar p}=D(1-e^{-t/t_c})^{2}.
\end{equation}
\vspace{28pc}

\section*{Appendix B: An example}

In Sect. IV we have derived our main result, 
Eq.(37), depending on the phenomenological quantities 
$C$ and $D$ through the relations ${\cal A}$ and ${\cal B}$ 
[see Eq.(31)]. It would be pretty catchy whether we could {\it a priori}
calculate them by means of the environmental  physics, that is, 
employing a theory of nonequilibrium thermodynamics. Unfortunately, 
at the moment there is no such theory. Yet, we can envisage   
the following physical situation: 
We imagine an environment having
$I_1$, $I_2$, and $I_3$ as subsystems, such that
within $I_1$ the Brownian particle is described by our
Fokker--Planck equation (12), while within $I_2$ it is
described by the Kramers equation (15), and within
$I_3$ its dynamics turns out to be governed by the
Smoluchowski equation (16). As outlined in Sect.III, supposing the equilibrium
Maxwell--Boltzmann distribution (14) to be a boundary condition
in the regions $I_2$ and $I_3$, we determine from (15) the
phase-space diffusion coefficient $A_{\bar p \bar p}=2\gamma \kappa_{B} T$,
and from (16) the $x$-space diffusion coefficient
$A_{xx}=2 \kappa_{B} T/\gamma$. Owing to the constraint (13)
we find $A_{x\bar p}=2 \kappa_{B} T$. [In contrast with $A_{xx}$ and 
$A_{\bar p \bar p}$, it is interesting to note that  the
phase-space diffusion coefficient $A_{x\bar p}$ is independent of the friction
constant $\gamma$. That absence of a dissipation-fluctuation relation indicates 
a nonequilibrium physical situation]. Therefore, as a whole our
medium $I=I_1+I_2+I_3$ could be considered as a thermal reservoir away from
the equilibrium  but with locally defined
temperature concept. That Brownian particle is then characterized by our  
non-Markovian Langevin equations  
\begin{equation}
\frac{d\bar P}{dt}= -\frac{dV}{dX}-\frac{\gamma}{m}\bar P+
\left[\ 2t_c \gamma \kappa_B T(1-e^{-t/t_c}) \right]^{1/3}\Psi,
\end{equation}
\begin{equation}
 \frac{dX}{dt}=\frac{\bar P}{m}+
 \left[\ \frac{2t_c  \kappa_B T}{\gamma^2}
 (1-e^{-t/t_c}) \right]^{1/3}\Psi,
\end{equation}
and by the corresponding Fokker--Planck equation (12) in the form 
\begin{equation}
\frac{\partial {\cal F}}{\partial t}=-\frac{\partial (A_x{\cal
F})}{\partial x}- \frac{\partial (A_{\bar p}{\cal F})}{\partial
\bar p}+\frac{ \kappa_B T}{\gamma} \frac{\partial^2 {\cal F}}{\partial
x^2}+ 2\kappa_B T\frac{\partial^2 {\cal F}}{\partial x\partial
\bar p}
 +\gamma \kappa_B T\frac{\partial^2 {\cal F}}{\partial \bar p^2},
\end{equation}
with
\begin{equation}
A_x=\frac{\bar p}{m}+ \left[\ \frac{(2\kappa_B T)^2}{t_c \gamma^4
(1-e^{-t/t_c})}\right]^{1/3} e^{-t/t_c},
\end{equation}
\begin{equation}
A_{\bar p}=-kx -\frac{\gamma \bar p}{m}+
\left[\ \frac{(2\kappa_B T)^2}{t_c \gamma
(1-e^{-t/t_c})}\right]^{1/3} e^{-t/t_c}.
\end{equation}
\vspace{1pc}

In the nonequilibrium
region the Kramers rate is given by (37) with $\beta=1/\kappa_B T$
and
\begin{equation}
{\cal A}=\frac{2(a-\gamma)}{m\left(\ a\sqrt{2\kappa_B T/\gamma}+
\sqrt{2\gamma \kappa_B T}\right)^2},
\end{equation}
\begin{equation}
{\cal B}=\frac{{\cal A}m}{(a-\gamma)}\frac{(2\kappa_B T)^{2/3}}{(\gamma t_c)^{1/3}}
e^{-\Delta \tau/t_c}\left[\ 1+ \frac{a(1-e^{-\Delta \tau/t_c})}{\gamma}\right]
\left(\ 1-e^{-\Delta \tau/t_c}\right)^{2/3}.
\end{equation}
In the case $\gamma \gg 2m \omega_b$ and $a \approx \gamma
+(m^2\omega^2_b/\gamma)$, the rate reads 
\begin{equation}
\Gamma=\frac{m\omega_a \omega_b}{4\pi \gamma}e^{-(1/4m\omega^2_b)
[(2\kappa_B T)/(\gamma t_c)^2]^{1/3}u(\Delta \tau)}
e^{-V(x_b)/\kappa_B T}
\end{equation}
with
\begin{equation}
u(\Delta \tau)=(2-e^{-\Delta \tau/t_c})^2(1-e^{-\Delta \tau/t_c})^{4/3}
e^{-2\Delta \tau/t_c}.
\end{equation}
We note that (77) may be written as
\begin{equation}
\Gamma_{\mbox {neq}}=\frac{1}{2}e^{-(1/4m\omega^2_b)
[(2\kappa_B T)/(\gamma t_c)^2]^{1/3}u(\Delta \tau)}\Gamma_{\mbox {eq}}
\end{equation}
relating the nonequilibrium rate $\Gamma_{\mbox {neq}}$ to the equilibrium one
\begin{equation}
\Gamma_{\mbox {eq}}=\frac{m\omega_a \omega_b}{2\pi\gamma}e^{-V(x_b)/\kappa_B T}
\end{equation}
which is obtained from (29) for $\gamma \gg 2m \omega_b$. 
In the Markovian limit, $\Delta \tau \gg t_c$, the nonequilibrium
rate (79) reduces to $\Gamma_{\mbox {neq}}=\Gamma_{\mbox
{eq}}/2$.\vspace{1pc}

On the other hand, for the low
friction  case, $\gamma \ll m \omega_b$, we obtain
\begin{equation}
\Gamma'_{{\mbox {neq}}}=\sqrt{\frac{\gamma}{m\omega_b}}
e^{-(1/m^2\omega^3_b)[2\kappa_B T\gamma/t_c^2]^{1/3}v(\Delta \tau)}
\Gamma'_{\mbox {eq}}
\end{equation}
with
\begin{equation}
v(\Delta \tau)=\left[\ 1+\left(\ \frac{1}{2}+\frac{m\omega_b}{\gamma}
\right)(1-e^{-\Delta \tau/t_c} ) \right]^{2}
( 1- e^{-\Delta \tau/t_c})^{4/3}e^{-2\Delta \tau/t_c}.
\end{equation}
In Eq.(81), $\Gamma'_{\mbox {eq}}=(\omega_a/\pi)e^{-V(x_b)/\kappa_B T}$ 
 is obtained from (29) as $\gamma \rightarrow 0$. 
In the Markovian limit, $\Delta \tau \gg t_c$, it follows that
$\Gamma'_{{\mbox {neq}}}=(\gamma/m\omega_b)^{1/2}\Gamma'_{\mbox {eq}}$.
\vspace{1pc} 

On the basis of our stochastic model for a non-Markovian Brownian particle  out  
equilibrium, from Eqs. (79) and (81) we draw the conclusion that the performance 
of the nonequilibrium escape rate is too small compared to 
the equilibrium situation. 
\vspace{39pc}



\end{document}